\documentclass[aps,prb,twocolumn,superscriptaddress]{revtex4}
\usepackage[latin1]{inputenc}
\usepackage[dvips]{graphicx}
\usepackage{amssymb,amsfonts,amsmath}

\begin{document}


\title{Two-dimensional electron liquid state at LaAlO$_3$-SrTiO$_3$ interfaces}

\author{M. Breitschaft}
\affiliation{Center for Electronic Correlations and Magnetism, University of Augsburg, D-86135 Augsburg, Germany}
\author{V. Tinkl}
\affiliation{Center for Electronic Correlations and Magnetism, University of Augsburg, D-86135 Augsburg, Germany}
\author{N. Pavlenko}
\affiliation{Center for Electronic Correlations and Magnetism, University of Augsburg, D-86135 Augsburg, Germany}
\affiliation{Institute for Condensed Matter Physics, 1 Svientsitsky Str., UA-79011 Lviv, Ukraine}
\author{S. Paetel}
\affiliation{Center for Electronic Correlations and Magnetism, University of Augsburg, D-86135 Augsburg, Germany}
\author{C. Richter}
\affiliation{Center for Electronic Correlations and Magnetism, University of Augsburg, D-86135 Augsburg, Germany}
\author{J. R. Kirtley}
\affiliation{Center for Electronic Correlations and Magnetism, University of Augsburg, D-86135 Augsburg, Germany}
\author{Y. C. Liao}
\affiliation{Center for Electronic Correlations and Magnetism, University of Augsburg, D-86135 Augsburg, Germany}
\author{G. Hammerl}
\affiliation{Center for Electronic Correlations and Magnetism, University of Augsburg, D-86135 Augsburg, Germany}
\author{V. Eyert}
\affiliation{Center for Electronic Correlations and Magnetism, University of Augsburg, D-86135 Augsburg, Germany}
\author{T. Kopp}
\affiliation{Center for Electronic Correlations and Magnetism, University of Augsburg, D-86135 Augsburg, Germany}
\author{J. Mannhart}
\email[Electronic address: ]{jochen.mannhart@physik.uni-augsburg.de}
\affiliation{Center for Electronic Correlations and Magnetism, University of Augsburg, D-86135 Augsburg, Germany}

\date{\today}

\begin{abstract}
Using tunneling spectroscopy we have measured the spectral density of states of the mobile, two-dimensional electron system generated at the LaAlO$_3$-SrTiO$_3$ interface. As shown by the density of states the interface electron system differs qualitatively, first, from the electron systems of the materials defining the interface and, second, from the two-dimensional electron gases formed at interfaces between conventional semiconductors.
\end{abstract}


\maketitle

Two-dimensional (2D) conducting electron systems are generated at interfaces between a large variety of insulating oxides \cite{Ohtomo2004,Mannhart2008}. These interfaces show a broad spectrum of different properties. The quantum-Hall effect has been found \cite{Tsukazaki2007}, for example, for the electron system at the ZnO-(Mg$_x$Zn$_{1-x}$O) interface. For the interface between LaTiO$_3$ and SrTiO$_3$, first explored experimentally by Ohtomo and Hwang \cite{Ohtomo2002}, spin ordering and ferro-orbital ordering has been predicted \cite{Okamoto06}. The most widely investigated electron system at oxide interfaces is the metallic state created at the interface between the charge transfer insulators LaAlO$_3$ and TiO$_2$-terminated SrTiO$_3$ \cite{Ohtomo2004}. This electron system forms a 2D superconductor with a $T_{\text{c}}$ of $\approx 250$\,mK that is easily tunable by electric gate fields \cite{Caviglia2008}. The LaAlO$_3$-SrTiO$_3$ interface has also been predicted \cite{Pentcheva2006} and reported \cite{Brinkman2007} to develop magnetic order. 

While several theoretical models have been developed to describe the electronic properties of these interfaces \cite{Pavlenko09,Pentcheva09,Schwingenschloegl2009,Schwingenschloegl2009a}, less information on the electronic structure has been provided by experiments. At room temperature its thickness has been inferred from scanning tunneling microscopy (STM) writing experiments \cite{Cen2008}, from photoemission \cite{Sing2009}, and from cross-sectional STM (Ref.\ \onlinecite{Basletic2008}) to be at most a few nanometers. Further, hard x-ray photoelectron emission has shown that the charge carriers at the interface occupy Ti\,$3d$ states \cite{Sing2009}. Studies of x-ray absorption spectroscopy furthermore revealed that energetically the crystal-field split Ti levels are rearranged, such that the $3d_{xy}$ levels are the first available states for the conducting electrons \cite{Salluzzo2009}.  

The spectral density of states (DOS) at the interface is a fundamental property that characterizes the electron system. As it furthermore can be calculated as well as measured, it is a key property for the understanding of the electron system at the interface. For measurements of the spectral DOS scanning tunneling spectroscopy (STS) is a powerful technique \cite{Feenstra1994a}, which has been used extensively to characterize 2D electron gases (2DEGs) in semiconductor systems. STS was employed, in particular, to probe surfaces of semiconducting thin films where electrons are confined by the film thickness \cite{Perraud2008}. STS was also used successfully to analyze cross-sectional cleavage planes of semiconductor heterostructures \cite{Salemink1989,Suzuki2007}. In addition, semiconductor surfaces, below which electrons are confined in band bending regions induced by ion implantation \cite{Wolovelsky1998} or surfaces at which electron gases were generated by adsorbates \cite{Morgenstern2003}, were explored.

Here we report on STS measurements of the spectral DOS of the electron system at the LaAlO$_3$-SrTiO$_3$ interface. We find the measured DOS to be in excellent agreement with the interface DOS calculated in density-functional theory (DFT), providing evidence that the tunneling current in the STS measurements is carried by interface states. The measured spectrum of the interface DOS and therefore the electron system differs qualitatively from the DOS of doped bulk SrTiO$_3$ or LaAlO$_3$. The electron system cannot be accurately described as a thin layer of doped SrTiO$_3$. The measurements reveal furthermore that the electron system also differs qualitatively from the hitherto known 2D electron systems at interfaces between conventional semiconductors. We find the electrons confined in multiple layers of quantum wells given by the ionic potentials of the TiO$_6$ octahedra. In these wells the electrons are subject to the correlations characteristic of the $d$ orbitals of the Ti ions. The spectral DOS is not a step function as is the case for standard semiconductor interfaces but rather resembles the DOS of Ti\,$3d$ states. Quantum wells and electronic systems of this kind are unknown from the 2DEGs in conventional semiconductors, in graphene or in ZnO. 

\begin{figure}
\begin{center}
\includegraphics[width=\columnwidth]{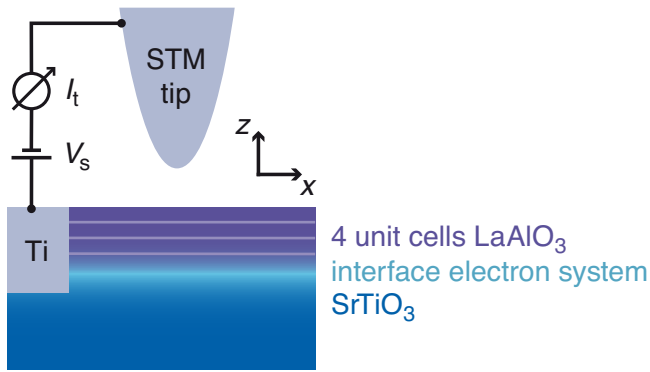}
\end{center}
\caption{Illustration of the experimental configuration. A metallic state is formed at the interface between the SrTiO$_3$ substrate and a 4 uc thick layer of LaAlO$_3$. Scanning tunneling microscopy and spectroscopy are  performed by monitoring the tunneling current $I_{\mathrm{t}}$ between the tip and the sample as a function of $V_{\mathrm{s}}$, the voltage at the sample relative to the tip.}
\label{fig1}
\end{figure}

For the studies, we fabricated LaAlO$_3$-SrTiO$_3$ heterostructures with 4 unit-cell (uc) thick ($\approx 1.6$\,nm) epitaxial LaAlO$_3$ layers to obtain measurable tunneling currents.  This thickness was chosen because it is the minimum thickness required to generate the conducting interface \cite{Thiel2006}. For larger LaAlO$_3$ thicknesses the tunneling current densities become impractically small. The samples were grown by standard pulsed laser deposition as described in Ref.\ \onlinecite{epaps}. For deposition the SrTiO$_3$ substrates were heated to 780$\,^{\circ}$C in an oxygen background pressure of $8\times 10^{-5}$\,mbar. The LaAlO$_3$ film growth was monitored by reflection high-energy electron diffraction. While SrTiO$_3$ surfaces are known to show numerous surface reconstructions \cite{Herger2007,Lu2010,Deak2007,Eglitis2008,Wang2009,Silly2006,Kubo2003,Newell2007}, x-ray diffraction showed no evidence of distortions of the LaAlO$_3$ films, which could be attributed to a SrTiO$_3$ surface reconstruction, suggesting that the LaAlO$_3$ growth stabilizes the standard SrTiO$_3$ structure at the interface. Titanium plugs filling ion etched holes were used to contact the interfaces. After a heating procedure in a preparation chamber \cite{epaps}, the samples were transferred \emph{in situ} to the scanning probe microscope (SPM), which operates in ultrahigh vacuum at $4.7$\,K. An iridium spall attached to a cantilever based on a quartz tuning fork \cite{Giessibl1998} with a spring constant of 1800\,N$/$m was used as a tip. The tip was treated \emph{in situ} by field emission \cite{epaps}. The cantilever was not excited mechanically during STM and STS measurements. The experimental setup is sketched in Fig.\ \ref{fig1}. Typical measurement parameters were tunneling currents of 10\,pA, sweep rates of $0.01$\,V$/$s and scanning speeds of 10\,nm$/$s.

\begin{figure}
\begin{center}
\includegraphics[width=\columnwidth]{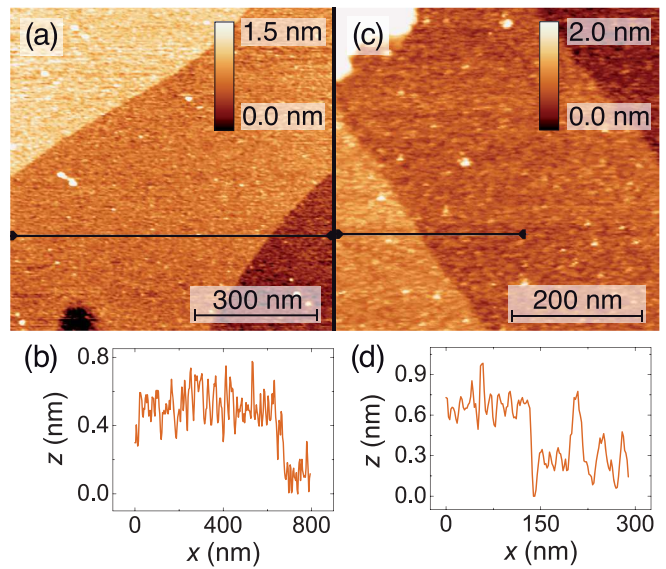}
\end{center}
\caption{Scanning probe microscopy images of LaAlO$_3$-SrTiO$_3$ heterostructures. (a)~Topographic FM-SFM image of the LaAlO$_3$ film. (b)~Profile taken along the line indicated in (a). (c)~Topographic STM image acquired on the LaAlO$_3$-SrTiO$_3$ heterostructure recorded with a scanning speed of 5\,nm$/$s, a bias voltage $V_{\mathrm{s}}=2$\,V, and a current set point of 10\,pA. (d)~Profile taken along the line indicated in (c). All data were taken at $4.7$\,K. For further detail see Ref.\ \onlinecite{epaps}.}
\label{fig2}
\end{figure}

Imaging the LaAlO$_3$-SrTiO$_3$ heterostructures by frequency modulation scanning force microscopy \cite{Albrecht1991} (FM-SFM) as well as by constant current STM revealed the standard step-and-terrace structure resulting from the slight miscut of the SrTiO$_3$ substrates (Fig.\ \ref{fig2}). While on more conventional samples excellent resolution was achieved with the SPM employed \cite{Hembacher2004}, it was impossible to obtain atomic resolution on the LaAlO$_3$-SrTiO$_3$ heterostructures \cite{epaps}.

\begin{figure}
\begin{center}
\includegraphics[width=62mm]{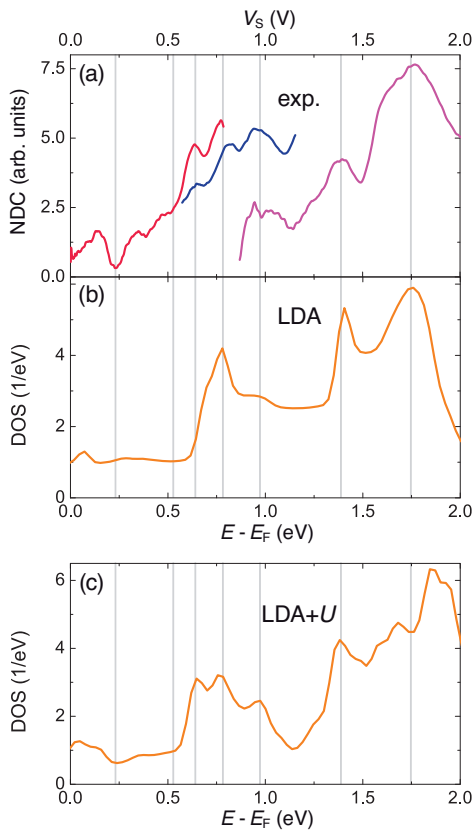}
\end{center}
\caption{Comparison of the measured differential conductance and DFT calculated state densities. (a)~$\mathrm{NDC}(V_{\mathrm{s}})=(\partial I_{\mathrm{t}}/\partial V_{\mathrm{s}})/(I_{\mathrm{t}}/V_{\mathrm{s}}+\epsilon)(V_{\mathrm{s}})$ characteristics with $\epsilon=1$\,pA$/$V measured at several sites located far away from topographic steps. The measurements were performed at $4.7$\,K with fixed tip-sample separations. The different colors reflect different tip-sample separations (purple: $I_{\text{t{,}stab}}=70$\,pA, $V_{\text{s{,}stab}}=2.4$\,V; blue: $I_{\text{t{,}stab}}=12$\,pA, $V_{\text{s{,}stab}}=1$\,V; red: $I_{\text{t{,}stab}}=12$\,pA, $V_{\text{s{,}stab}}=0.8$\,V). The data were averaged over an interval of 75\,mV. The purple characteristic was measured on a different sample than the red and blue ones. (b)~Ti\,$3d$ DOS of the interface TiO$_2$ layer calculated using LDA for a supercell with a 4 uc thick LaAlO$_3$ layer. (c)~Ti\,$3d$ DOS of the interface TiO$_2$ layer calculated using LDA$+U$. In (a)--(c) the positions of characteristic features in the NDC are marked with gray lines.}
\label{fig3}
\end{figure}

Conductance-voltage characteristics $\partial I_{\mathrm{t}}/\partial V_{\mathrm{s}}(V_{\mathrm{s}})$ were measured using a standard lock-in technique \cite{epaps}. Simultaneously, the tunneling current was measured as a function of voltage. The normalized differential conductance $\mathrm{NDC}\equiv (\partial I_{\mathrm{t}}/\partial V_{\mathrm{s}})/(I_{\mathrm{t}}/V_{\mathrm{s}}+\epsilon)$ was determined as a measure of the sample DOS \cite{Feenstra1987}. The spectra were taken on sample areas where the step-and-terrace structure was resolved in STM topography. We found the characteristic spectroscopic features to be reproducible across four samples \cite{epaps}. Figure \ref{fig3}(a) shows a representative dependence of the NDC on voltage. The conductances are minute for negative voltages (tunneling from occupied sample states). For positive voltages (tunneling into unoccupied sample states) the spectroscopically accessible energy range is limited at low voltages by small tunneling conductances and at high voltages by large electric fields destabilizing the tunneling gap. To measure the tunneling characteristics at a given sample location over a large voltage range, several spectra were therefore taken at different tip-sample separations as determined according to the tunneling currents $I_{\text{t{,}stab}}$ at given gap voltages $V_{\text{s{,}stab}}$. Three characteristics measured with different tip-sample separations, from two different samples, are shown in Fig.\ \ref{fig3}(a). In these spectra, clear peaks are seen at $\approx 0.6$, $\approx 0.8$, $\approx 1$, $\approx 1.4$, and $\approx 1.8$\,V.

To identify the electron states carrying the measured tunneling current and to explore the role of electronic correlations at the interface we compare the measured DOS to predictions of DFT. We performed local-density approximation (LDA) and LDA$+U$ calculations \cite{wien2k,Anisimov93} of the layer-resolved DOS of LaAlO$_3$-SrTiO$_3$ heterostructures. Further information on these calculations is given in Ref.\ \onlinecite{epaps}. While differences are present in details, the calculated state densities and the effective electron mass of $3.25$ bare electron masses are consistent with those reported in Refs.\ \onlinecite{Schuster08,Pentcheva09,Pavlenko09}, and \onlinecite{Copie2009}.

In Fig.\ \ref{fig3}(b) the Ti\,$3d$ DOS of the interface TiO$_2$ layer calculated using LDA for a supercell with a 4 uc thick LaAlO$_3$ layer on SrTiO$_3$ is shown. According to the calculation, electronic reconstruction leads to a doping of O\,$2p$ states located in the topmost AlO$_2$ layer with holes and of Ti\,$3d\,t_{\mathrm{2g}}$ states located at the interface with electrons. Experimentally, it is only the interface which is found to be conducting. In the total DOS between 0 and 2\,eV the Ti\,$3d\,t_{\mathrm{2g}}$ orbitals located in the interfacial TiO$_2$ layer prevail. The other, small, contributions are provided by the TiO$_2$ planes of adjacent SrTiO$_3$ layers and, below $\approx 0.5$\,eV, by the O\,$2p$ states of the surface. The Ti\,$3d\,e_{\mathrm{g}}$ states contribute at energies above $\approx 2.8$\,eV and the La\,$5d$ states at energies above $\approx 2.2$\,eV. The measured peaks at $\approx 0.8$, $\approx 1.4$, and $\approx 1.8$\,V are also present in the calculated DOS. The measured peaks at $\approx 0.6$ and $\approx 1$\,V, however, are not represented in the LDA result.

The LDA$+U$ calculations of the interface electron system consider an on-site Coulomb repulsion $U=2$\,eV and a Hund coupling $J=0.8$\,eV in the Ti\,$3d$ shell. The choice of conservative values for $U$ and $J$ does not imply these values characterize the system best. Figure \ref{fig3}(c) shows the DOS calculated for the supercell using LDA$+U$. The DOS exhibits additional peaks at $\approx 0.6$ and $\approx 1$\,eV, generated by the splitting of the Ti\,$3d_{xz}+3d_{yz}$ bands due to the interorbital interactions caused by the finite $U$ and $J$. Remarkably, these peaks are observed experimentally but are missing in the LDA DOS.

We note that the experimental hump at $\approx 1.8$\,eV is broader than the corresponding structure in LDA. However, LDA$+U$ generates a structure of approximately the measured width but with finer structures. These fine structures reflect the formation of the upper Hubbard bands, which is a fundamental effect of correlated electron systems, arising when $U$ is on the order of the bandwidth or larger. Indeed, the calculated width of the Ti\,$3d\,t_{\mathrm{2g}}$ band is $\approx 2$\,eV$=U$.

The good agreement between experiment and calculation suggests that the electron states carrying the measured tunneling current are the ones calculated in DFT. For energies between $0.5$ and 2\,eV, mainly Ti\,$3d_{xz}+3d_{yz}$ and Ti\,$3d_{xy}$ orbitals of the interface TiO$_2$ layer contribute to the calculated DOS and the prominent peaks result from these orbitals; tunneling occurs into Ti\,$3d$ orbitals at the interface, the significant contributions arising from the Ti\,$3d\,t_{\mathrm{2g}}$ states. These results are consistent with the results of recent photoabsorption measurements \cite{Salluzzo2009}, from which it was concluded that the lowest unoccupied states are Ti\,3$d_{xy}$ states.

\begin{figure}
\begin{center}
\includegraphics[width=\columnwidth]{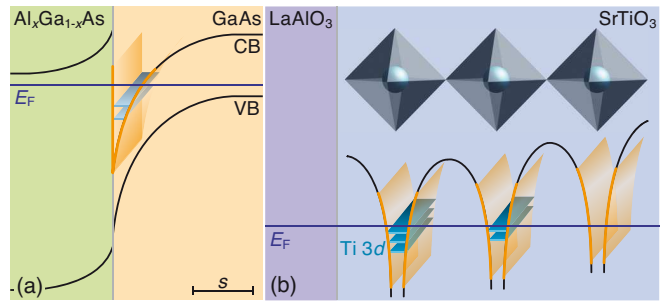}
\end{center}
\caption{Illustration of the configuration of 2D electron systems in standard semiconductor interfaces and at the LaAlO$_3$-SrTiO$_3$ interface. (a) At the interface between the semiconductors an electron gas is generated at a potential well created by band bending, which typically has a width of tens of nanometers as determined by the electronic screening length $s$. The electron states can be approximated by the states of free electrons in this potential well. (b) At the oxide interface the potential well is provided by the Coulomb potential of the titanium ions in the TiO$_6$ octahedra and, to a smaller extent, by band bending. These potential wells are narrower than those at semiconductor interfaces; the ``resonant'' electron states are well approximated by the Ti\,$3d\,t_{\mathrm{2g}}$ states, which form a 2D electron system extended parallel to the interface. Due to the electronic correlations of the oxide lattices, the mobile electrons form an electron liquid.}
\label{fig4}
\end{figure}

The fact that the experimental DOS is matched significantly better by the LDA$+U$ calculation than by the LDA calculation provides evidence that the electron system at the interface is correlated with substantial values of $U$ and $J$ on the Ti\,$3d$ orbitals. Therefore, this 2D electron system is a liquid. This electron liquid is formed by correlated electrons, which can move parallel to the interface, but are constrained in their perpendicular motion by the Coulomb potentials of the titanium ions of the final TiO$_2$ layers and also, to a smaller degree, by band bending (Fig.\ \ref{fig4}).

Interfaces in oxides therefore broaden the spectrum of available 2D electron systems from the 2DEGs of conventional semiconductors to also include systems with sizable electronic correlations. Such correlation effects, in combination with the already intriguing physics of 2D electron systems, promise unprecedented electronic phenomena. Influenced by electronic correlations generated in the ionic lattices of the oxides, electron systems at oxide interfaces have exceptional properties, possibly enabling devices with hitherto unknown characteristics.

We thank D.\ Bonn, R.\ Claessen, M.\ Fiebig, F.\ J.\ Giessibl, J.\ Repp, C.\ W.\ Schneider and D.\ Vollhardt for helpful discussions. This work was supported by the Deutsche Forschungsgemeinschaft (TRR 80) and by the European Union (OxIDes). J.\ R.\ Kirtley was supported by the Alexander von Humboldt foundation. Grants of computer time from the Leibniz-Rechenzentrum München through HLRB (h1181) and DEISA (OXSIM) are gratefully acknowledged.

\end{document}



\title{Two-dimensional electron liquid state at LaAlO$_3$-SrTiO$_3$ interfaces -- supplementary material}

\author{M. Breitschaft}
\affiliation{Center for Electronic Correlations and Magnetism, University of Augsburg, D-86135 Augsburg, Germany}
\author{V. Tinkl}
\affiliation{Center for Electronic Correlations and Magnetism, University of Augsburg, D-86135 Augsburg, Germany}
\author{N. Pavlenko}
\affiliation{Center for Electronic Correlations and Magnetism, University of Augsburg, D-86135 Augsburg, Germany}
\affiliation{Institute for Condensed Matter Physics, 1 Svientsitsky Str., UA-79011 Lviv, Ukraine}
\author{S. Paetel}
\affiliation{Center for Electronic Correlations and Magnetism, University of Augsburg, D-86135 Augsburg, Germany}
\author{C. Richter}
\affiliation{Center for Electronic Correlations and Magnetism, University of Augsburg, D-86135 Augsburg, Germany}
\author{J. R. Kirtley}
\affiliation{Center for Electronic Correlations and Magnetism, University of Augsburg, D-86135 Augsburg, Germany}
\author{Y. C. Liao}
\affiliation{Center for Electronic Correlations and Magnetism, University of Augsburg, D-86135 Augsburg, Germany}
\author{G. Hammerl}
\affiliation{Center for Electronic Correlations and Magnetism, University of Augsburg, D-86135 Augsburg, Germany}
\author{V. Eyert}
\affiliation{Center for Electronic Correlations and Magnetism, University of Augsburg, D-86135 Augsburg, Germany}
\author{T. Kopp}
\affiliation{Center for Electronic Correlations and Magnetism, University of Augsburg, D-86135 Augsburg, Germany}
\author{J. Mannhart}
\email[Electronic address: ]{jochen.mannhart@physik.uni-augsburg.de}
\affiliation{Center for Electronic Correlations and Magnetism, University of Augsburg, D-86135 Augsburg, Germany}



\maketitle

\subsection{Sample preparation}
The LaAlO$_3$ films were epitaxially grown by pulsed laser deposition on TiO$_2$-terminated $(001)$ surfaces \cite{Kawasaki1994,Koster1998} of SrTiO$_3$ single crystals. The deposition was conducted at a substrate temperature of 780$\,^{\circ}$C and at an oxygen background pressure of $8\times 10^{-5}$\,mbar. The film growth was monitored by reflection high-energy electron diffraction. The samples were cooled in 400\,mbar of oxygen. During the cooling the samples were held at 600$\,^{\circ}$C for one hour. After cool-down an aluminum shadow mask was applied to define a stripe along one edge of a sample. Within this stripe the SrTiO$_3$ substrates were laid bare by Ar ion etching and then covered by electron beam evaporation with titanium to provide a contact to the interface electron system.

\subsection{Sample treatment by heating}
To achieve reproducible surface conditions after the samples were transferred in air to the preparation chamber of the SPM, they were radiatively heated to $\approx 400\,^{\circ}$C in an oxygen pressure of $10^{-2}$\,mbar (background pressure $\approx 5\times 10^{-7}$\,mbar).

\subsection{Tip treatment}
A tip treatment by field emission at gap voltages up to $V_{\mathrm{s}}=400$\,V and emission currents up to several microamperes, while the tip was placed in proximity to the LaAlO$_3$ surface, was necessary to obtain reproducible results in topographic STM imaging and STS. Usually, the field emission procedure damaged the sample locally. We do not expect this tip preparation technique to yield a clean metallic tip. The remaining contamination made it difficult to establish stable tunneling. The tunneling current $I_{\mathrm{t}}$ and the differential conductance $\partial I_{\mathrm{t}}/\partial V_{\mathrm{s}}$ tended to fluctuate on timescales from fractions of a second to several minutes, which we attribute to spontaneous changes of the tunneling gap configuration. By repeating the tip preparation procedure described, however, it was possible to reduce the fluctuations such that reproducible experimental results were obtained.

\subsection{FM-SFM imaging (Fig.\ 2(a) and (b))}
The image in Fig.\ 2(a) was recorded with a scanning speed of 16\,nm$/$s and a bias voltage of $V_{\mathrm{s}}=0.1$\,V. The eigenfrequency of the cantilever was $f_0=24296.8$\,Hz, the quality factor $Q=17832$ and the set oscillation amplitude $A=1.8$\,\AA. The step heights in Fig.\ 2(a) were used to calibrate the scanner of the SPM. As shown by Fig.\ 2(a) and (b), in FM-SFM the surfaces are flat with an apparent roughness of $0.7$\,\AA$_{\mathrm{rms}}$. This value does not characterize the surface morphology quantitatively, because dissipation of vibrational energy of the cantilever may have caused topographic artifacts.

\subsection{STM imaging (Fig.\ 2(c) and (d))}
We attribute the apparent surface roughness of 1\,\AA$_{\mathrm{rms}}$ seen in STM topography (Fig.\ 2(c) and (d)) mainly to local variations of the tunneling barrier height, likely induced by adsorbates.

\subsection{STS lock-in technique}
The differential conductance $\partial I_{\mathrm{t}}/\partial V_{\mathrm{s}}$ was measured by adding to the bias voltage $V_{\mathrm{s}}$ an AC voltage with an amplitude $\hat{V}_{\mathrm{mod}}$ of several millivolts and a frequency $f_{\mathrm{mod}}$ on the order of 100\,Hz. The resulting in-phase AC component of the current signal was measured using lock-in detection. The ratio of its amplitude $\Delta I_{\mathrm{t}}$ and $\hat{V}_{\mathrm{mod}}$ approximates the differential conductance, $\Delta I_{\mathrm{t}}/\hat{V}_{\mathrm{mod}}\approx\partial I_{\mathrm{t}}/\partial V_{\mathrm{s}}$. Differences between this measured differential conductance and numerical derivatives of current-voltage characteristics were found to be insignificant. Tunneling current and differential conductance were recorded simultaneously with an acquisition time of 120\,s per sweep direction.

\subsection{STS parameters in Fig.\ 3(a)}
All spectra in Fig.\ 3(a) were measured with $\hat{V}_{\mathrm{mod}}=5\,\mathrm{mV_{rms}}$ and $f_{\mathrm{mod}}=77$\,Hz.

\subsection{Reproducibility of STS measurements}
Contaminants located between the tip and the LaAlO$_3$ surface may possibly distort the spectroscopic results. To understand the role of contaminants we investigated four samples. We regularly obtained stable tunneling for gap voltages up to $V_{\mathrm{s}}=1.2$\,V and observed the three characteristic peaks between $V_{\mathrm{s}}=0.5$\,V and $V_{\mathrm{s}}=1.2$\,V shown by Fig.\ 3(a). As we must expect that in these studies the configurations of adsorbates in the tunneling barrier were not identical from experiment to experiment, we conclude that the spectroscopic features cannot be caused by contaminants.

However, adsorbates may induce fluctuations frequently observed in the tunneling current $I_{\mathrm{t}}$ and the differential conductance $\partial I_{\mathrm{t}}/\partial V_{\mathrm{s}}$. The spectroscopic resolution was limited in most cases by such fluctuations. As explained earlier, a tip treatment by field emission made it possible to obtain stable tunneling, yielding reproducible spectra with low noise and low hysteresis.

\begin{figure}
\begin{center}
\includegraphics[width=90mm]{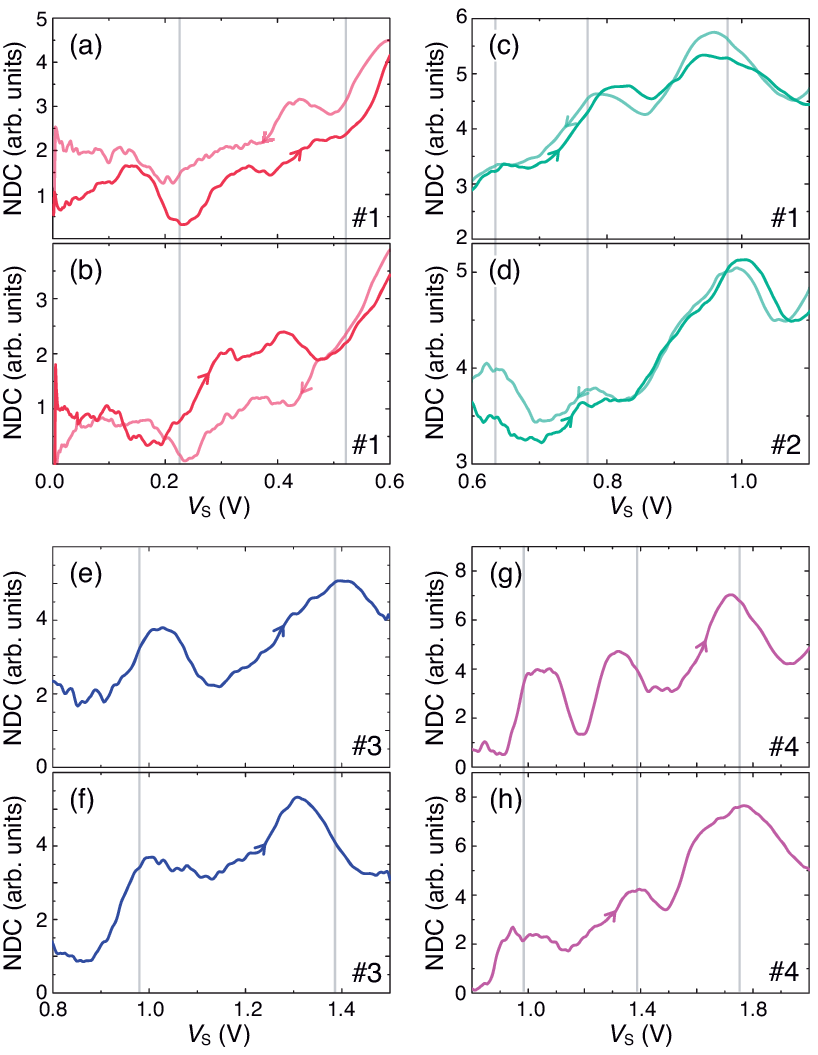}
\end{center}
\caption{$\mathrm{NDC}(V_{\mathrm{s}})=(\partial I_{\mathrm{t}}/\partial V_{\mathrm{s}})/(I_{\mathrm{t}}/V_{\mathrm{s}}+\epsilon)(V_{\mathrm{s}})$ characteristics with $\epsilon=1$\,pA$/$V acquired on samples 1 -- 4. The spectra were measured at $4.7$\,K with fixed tip-sample separations at sites far away from topographic steps. The different colors reflect different tip-sample separations. (a)\&(b)~Stabilization parameters: $V_{\mathrm{s{,}stab}}=0.8$\,V, $I_{\mathrm{t{,}stab}}=12$\,pA; lock-in parameters: $\hat{V}_{\mathrm{mod}}=5\,\mathrm{mV_{rms}}$, $f_{\mathrm{mod}}=77$\,Hz. (c)~Stabilization parameters: $V_{\mathrm{s{,}stab}}=1$\,V, $I_{\mathrm{t{,}stab}}=12$\,pA; lock-in parameters: $\hat{V}_{\mathrm{mod}}=5\,\mathrm{mV_{rms}}$, $f_{\mathrm{mod}}=77$\,Hz. (d)~Stabilization parameters: $V_{\mathrm{s{,}stab}}=1$\,V, $I_{\mathrm{t{,}stab}}=12$\,pA; lock-in parameters: $\hat{V}_{\mathrm{mod}}=10\,\mathrm{mV_{rms}}$, $f_{\mathrm{mod}}=161$\,Hz \cite{not_heated}. (e)\&(f)~Stabilization parameters: $V_{\mathrm{s{,}stab}}=1.5$\,V, $I_{\mathrm{t{,}stab}}=30$\,pA; lock-in parameters: $\hat{V}_{\mathrm{mod}}=2.83\,\mathrm{mV_{rms}}$, $f_{\mathrm{mod}}=977$\,Hz. (g)\&(h)~Stabilization parameters: $V_{\mathrm{s{,}stab}}=2.4$\,V, $I_{\mathrm{t{,}stab}}=70$\,pA; lock-in parameters: $\hat{V}_{\mathrm{mod}}=5\,\mathrm{mV_{rms}}$, $f_{\mathrm{mod}}=77$\,Hz. The characteristics in (g) and (h) were measured at the same sample site. The spectrum shown in (a) is also plotted in Fig.\ 3(a) with a red line, the spectrum shown in (c) is also plotted in Fig.\ 3(a) with a blue line, the spectrum shown in (h) is also plotted in Fig.\ 3(a) with a purple line. The data were averaged over an interval of 75\,mV. As a guide to the eye, gray lines are drawn at the same positions as in Fig.\ 3(a).}
\label{figS1}
\end{figure}

As shown in Fig.\ \ref{figS1}, the characteristic features of the spectra (Fig.\ 3(a)) were reproducible across several samples. In Fig.\ \ref{figS1}(g) and (h) characteristics of the NDC are shown, which were measured at the same sample site and at nominally equal tip-sample distances. After the acquisition of the spectrum in Fig.\ \ref{figS1}(g), the tunneling gap configuration apparently changed. The measurements following the acquisition of the characteristic in Fig.\ \ref{figS1}(g) were accompanied by an increasing destabilization of the tunneling conditions, resulting in relatively large fluctuations of the tunneling current and differential conductance. The spectroscopic peak structure was partially obscured. During further measurements the fluctuations of the tunneling current and differential conductance decreased, which we attribute to the formation of a new, stable tunneling gap configuration. The characteristic shown in Fig.\ \ref{figS1}(h) measured with this gap configuration differs in overall shape from the characteristic in Fig.\ \ref{figS1}(g). In both spectra, however, the characteristic spectroscopic features appear at almost the same voltages, independent of the tunneling gap configuration.

For gap voltages $V_{\mathrm{s}}$ larger than $1.5$\,V stable tunneling was rarely obtained. Characteristics with a high resolution between $V_{\mathrm{s}}=1.5$\,V and $V_{\mathrm{s}}=2$\,V as shown in Fig.\ 3(a) and Fig.\ \ref{figS1}(g),(h) were obtained during one measurement session only.

All tunneling spectra shown in this work were acquired at moderate tip-sample distances, avoiding strong repulsive force interactions between the tip and the sample which could possibly induce mechanical strain on the sample. Within this distance regime, tunneling conductances became minute at gap voltages below $0.5$\,V and the signal-to-noise ratio of the tunneling current became small. Occasionally, it was nevertheless possible to derive the sample DOS for energies between 0\,eV and $0.5$\,eV. Figure \ref{figS1}(a) and (b) compare two spectra where this seems to be the case. Albeit relatively high hysteresis, both NDC characteristics share distinct features.

\subsection{Possible contribution of O\,$2p$ states at energies below $0.5\,$eV}

\begin{figure}
\begin{center}
\includegraphics[width=70mm]{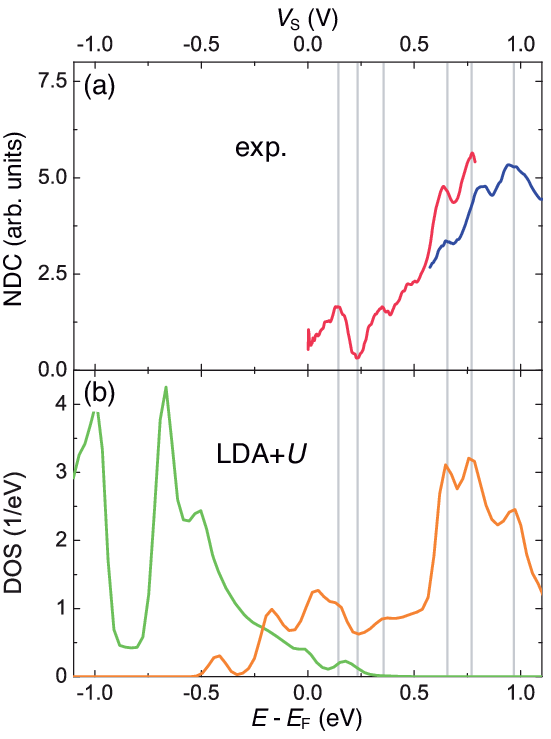}
\end{center}
\caption{Comparison of the measured differential conductance and DFT calculated state densities. (a)~$\mathrm{NDC}(V_{\mathrm{s}})=(\partial I_{\mathrm{t}}/\partial V_{\mathrm{s}})/(I_{\mathrm{t}}/V_{\mathrm{s}}+\epsilon)(V_{\mathrm{s}})$ characteristics with $\epsilon=1$\,pA$/$V measured at several sites located far away from topographic steps. The measurements were performed at $4.7$\,K with fixed tip-sample separations. The different colors reflect different tip-sample separations (blue: $I_{\mathrm{t{,}stab}}=12$\,pA, $V_{\mathrm{s{,}stab}}=1$\,V; red: $I_{\mathrm{t{,}stab}}=12$\,pA, $V_{\mathrm{s{,}stab}}=0.8$\,V). The data were averaged over an interval of 75\,mV. Both characteristics are also plotted in Fig.\ 3(a). (b)~Ti\,$3d$ DOS of the interface TiO$_2$ layer (orange) and O\,$2p$ DOS of the topmost AlO$_2$ layer adjoining vacuum (green) calculated using LDA$+U$ for a supercell with a 4 uc thick LaAlO$_3$ layer. In (a) and (b) the positions of characteristic features in the calculated Ti\,$3d$ DOS are marked with gray lines.}
\label{figS2}
\end{figure}

According to the DFT calculations, between $0.5\,$eV and 2\,eV Ti\,$3d$ states dominate the DOS. For smaller energies also O\,$2p$ states of the topmost AlO$_2$ layer adjoining vacuum contribute and may play a role in tunneling. Figure \ref{figS2} compares the measured differential conductance NDC (Fig.\ \ref{figS2}(a)) with the O\,$2p$ DOS of the topmost AlO$_2$ layer and the Ti\,$3d$ DOS of the interface TiO$_2$ layer calculated using LDA$+U$ (Fig.\ \ref{figS2}(b)).

\subsection{LDA and LDA$+U$ calculations}
\begin{figure}
\begin{center}
\includegraphics[width=30mm]{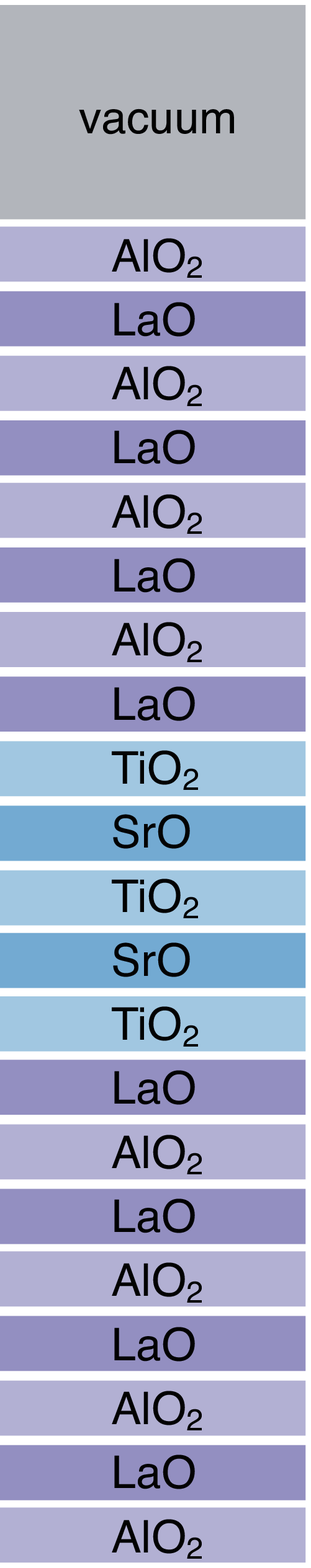}
\end{center}
\caption{Layer structure of the supercells used for the DFT calculations.}
\label{figS6}
\end{figure}

The spectral density of states of the LaAlO$_3$-SrTiO$_3$ bilayer was calculated for a supercell comprising 4 uc LaAlO$_3$, SrTiO$_3$ layers ($\approx 1$\,nm), 4 uc LaAlO$_3$ and vacuum ($\approx 1.3$\,nm), as sketched in Fig.\ \ref{figS6}. We employ the WIEN2k program package \cite{wien2k}, using 120 $\boldsymbol{k}$-points (15 $\boldsymbol{k}$-points in the irreducible wedge of the first Brillouin zone) for the LDA calculation, 600 $\boldsymbol{k}$-points (21 $\boldsymbol{k}$-points in the irreducible wedge of the first Brillouin zone), a Coulomb repulsion $U=2$\,eV and a Hund coupling $J=0.8$\,eV in the Ti\,$3d$ shell for the LDA$+U$ calculation \cite{Anisimov93}. To avoid a spurious mixing of the La\,$f$ states with the Ti\,$3d$ bands, a large $U$ of 8\,eV is imposed on the La\,$f$ states, a procedure which was introduced in Ref.\ \onlinecite{Okamoto06}. The calculations involved structural relaxation along all crystal axes.

\begin{figure}
\begin{center}
\includegraphics[width=70mm]{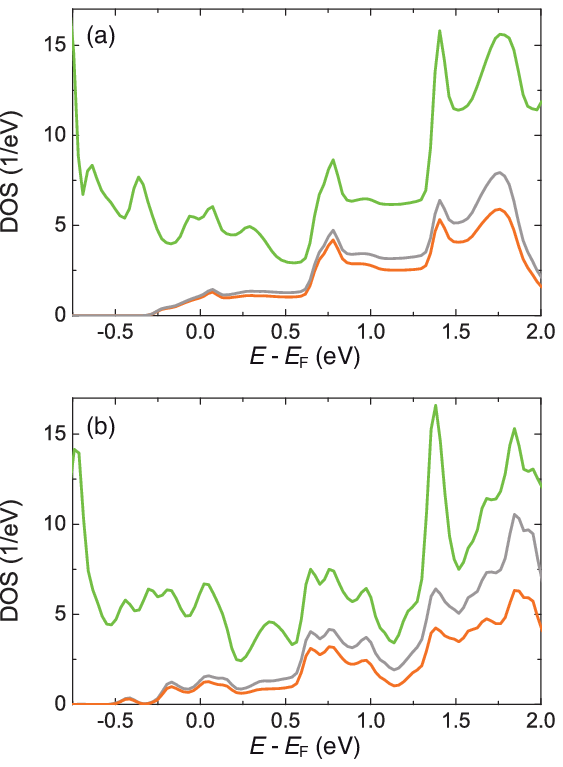}
\end{center}
\caption{Calculated total DOS of the full supercell with a 4 uc thick LaAlO$_3$ layer on SrTiO$_3$ and state densities of the interface TiO$_2$ layer. The total DOS of the full supercell (green), the DOS of the interface TiO$_2$ layer (gray) and the Ti\,$3d$ DOS of this layer (orange) are shown. Note that the partial DOS (gray and orange) arise from integration over non-overlapping muffin-tin spheres, which neglects the remaining, interstitial regions. Therefore, the sum of the partial DOS is smaller than the total DOS (green). (a)~State densities calculated using LDA. (b)~State densities calculated using LDA$+U$.}
\label{figS3}
\end{figure}

Figure \ref{figS3} shows the total DOS of the full supercell (green), the DOS of the interface TiO$_2$ layer (gray) and the Ti\,$3d$ DOS of the interface TiO$_2$ layer (orange), calculated both using LDA (Fig.\ \ref{figS3}(a)) and LDA+$U$ (Fig.\ \ref{figS3}(b)). Between $0.5$\,eV and 2\,eV the main contributions to the total DOS of the full supercell arise from Ti\,$3d\,t_{\mathrm{2g}}$ orbitals located in the interface TiO$_2$ layer. At energies below $\approx 0.5$\,eV, O\,$2p$ states located in the topmost AlO$_2$ layer of the supercell contribute to the total DOS.

While Ti\,$3d$ states of the interface TiO$_2$ layer almost completely dominate the calculated total DOS of the full supercell at energies between $0.5$\,eV and 2\,eV and generate the three peaks between $\approx 0.6$\,eV and $\approx 1$\,eV, contributions from Ti\,$3d$ states of the adjacent TiO$_2$ layer considerably heighten the peak at $1.4$\,eV. In the total DOS of the full supercell (see Fig.\ \ref{figS3}) this peak appears to have a larger weight than found experimentally (Fig.\ 3(a)), which is possibly related to the current tunneling predominantly into electron states of the final TiO$_2$ layer.

\begin{figure}
\begin{center}
\includegraphics[width=70mm]{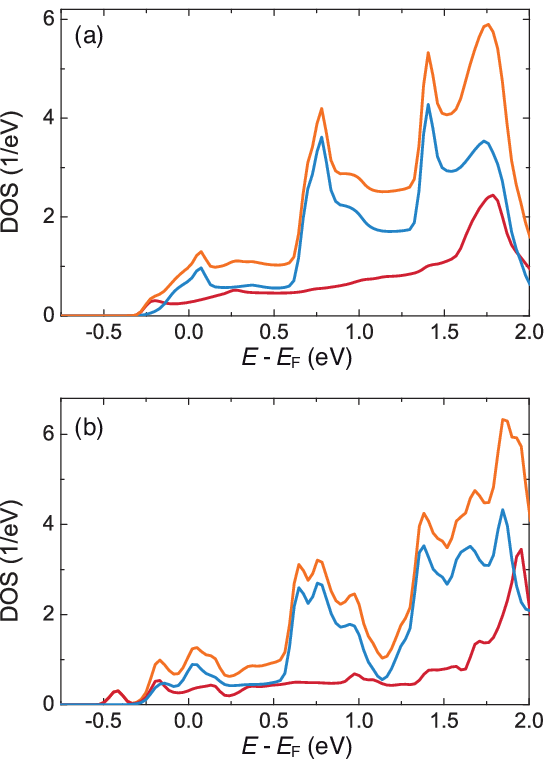}
\end{center}
\caption{Orbital contributions to DFT calculated Ti\,$3d$ state densities of the Ti ions in the interface TiO$_2$ layer. The total Ti\,$3d$ DOS (orange) contains contributions arising from Ti\,$3d_{xy}$ orbitals (red) and  Ti\,$3d_{xz}+3d_{yz}$ orbitals (blue). (a)~Ti\,$3d$ state densities calculated using LDA. (b)~Ti\,$3d$ state densities calculated using LDA$+U$.}
\label{figS4}
\end{figure}

Figure \ref{figS4} shows the orbital contributions to the Ti\,$3d$ DOS of the interface TiO$_2$ layer both for LDA (Fig.\,\ref{figS4}(a)) and LDA+$U$ (Fig.\,\ref{figS4}(b)).

\subsection{Calculated state densities of bulk LaAlO$_3$ and SrTiO$_3$}
\begin{figure}
\begin{center}
\includegraphics[width=70mm]{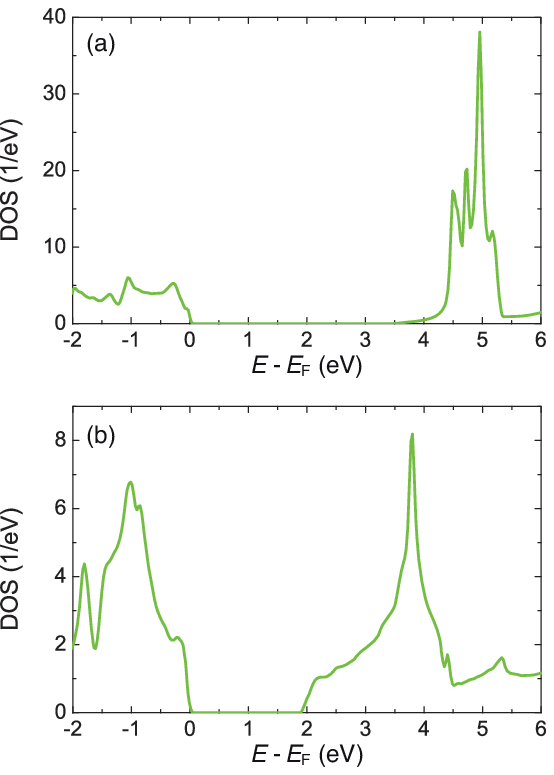}
\end{center}
\caption{State densities of bulk LaAlO$_3$ (a) and bulk SrTiO$_3$ (b) calculated using LDA.}
\label{figS5}
\end{figure}

Figure \ref{figS5} shows the total state densities of bulk LaAlO$_3$ (Fig.\,\ref{figS5}(a)) and of bulk SrTiO$_3$ (Fig.\,\ref{figS5}(b)) calculated using LDA. The state densities are calculated employing the WIEN2k program package, using 3000 $\boldsymbol{k}$-points (84 $\boldsymbol{k}$-points in the irreducible wedge of the first Brillouin zone) for the LaAlO$_3$ DOS and 3375 $\boldsymbol{k}$-points (120 $\boldsymbol{k}$-points in the irreducible wedge of the first Brillouin zone) for the SrTiO$_3$ DOS. The DOS of the LaAlO$_3$-SrTiO$_3$ heterostructure is different from the DOS of either constituent.

\clearpage

\clearpage

\clearpage

\clearpage

\clearpage

\clearpage

